\documentclass[12pt,a4paper,final]{iopart}

\usepackage{amsmath}
\usepackage{iopams}  
\usepackage{graphicx}
\usepackage{todonotes}
\usepackage{subfig}
\usepackage[breaklinks=true,colorlinks=true,linkcolor=blue,urlcolor=blue,citecolor=blue]{hyperref}

\begin{document}

\title[]{Numerical solution of the relativistic single-site scattering problem for the Coulomb and the Mathieu potential}

\author{Matthias Geilhufe$^{1}$, Steven Achilles$^{2}$, Markus Arthur K\"obis$^{3}$, Martin Arnold$^{3}$, Ingrid Mertig$^{1,2}$, Wolfram Hergert$^{2}$ and Arthur Ernst$^{1}$}
\address{$^{{1}}$Max Planck Institute of Microstructure Physics, Weinberg 2, D-06120 Halle (Saale), Germany}
\address{$^{{2}}$Martin Luther University Halle-Wittenberg, Institute of Physics, Von-Seckendorff-Platz 1, D-06120 Halle (Saale), Germany}
\address{$^{{3}}$Martin Luther University Halle-Wittenberg, Institute for Mathematics, Theodor-Lieser-Str. 5, D-06120 Halle (Saale), Germany }
\ead{geilhufe@mpi-halle.de}

\begin{abstract}
  For a reliable fully-relativistic Korringa-Kohn-Rostoker Green
  	function method, an accurate solution of the underlying single-site scattering
  problem is necessary. We present an extensive discussion on
  numerical solutions of the related differential equations by
  means of standard methods for a direct solution and by means of integral equations. Our
  implementation is tested and exemplarily demonstrated for a spherically symmetric treatment of a Coulomb potential and for a Mathieu potential to cover the full-potential implementation. For the Coulomb
  potential we include an analytic discussion of the asymptotic behaviour of irregular scattering solutions close to the origin ($r\ll1$).
\end{abstract}

\vspace{2pc}

\section{Introduction}
The Korringa-Kohn-Rostoker Green function method (KKR) in the
framework of density functional theory \cite{Hohenberg1964} has been a
powerful tool for electronic structure calculations in the past
decades.  Although it is based on the early work of Korringa
\cite{korringa1947} and Kohn and Rostoker \cite{kohn1954} in 1947 and
1954, respectively, an active development of the method is present
also today \cite{Ebert2011}. Originally, the KKR method was formulated
for non-relativistic approximations using muffin-tin spheres for the description of the potentials. Significant extensions of the original method are given by the relativistic KKR method \cite{feder1983,Strange1984}, by
the full-potential KKR method \cite{Drittler1991a,Drittler1991b} and
by the combination of both, the fully-relativistic full-potential KKR
method \cite{Huhne1998}. These developments allow for precise
calculations of total energies and forces including relativistic
effects. This opens new perspectives for many applications \cite{Ebert2011} such as structure optimisations, equations of states and phonons.

The KKR method is based on the multiple scattering theory, whereas a
periodic system is separated into disjoint atomic regions which
represent scattering centres. Since the Green function is constructed
from the regular and irregular single-site scattering solutions at
each lattice site, an accurate numerical solution of the underlying
differential equations is crucial. Within the non-relativistic
full-potential KKR method, it was suggested by Drittler
\cite{Drittler1991a}, to solve the underlying single-site scattering
problem by means of a Lippmann-Schwinger equation, i.e. in terms of
integral equations. In general, these integral equations can be solved
iteratively via Born series. Later it was suggested to reformulate the
original approach by means of integral equations of Volterra type
to achieve better convergence properties
\cite{zeller2004,Ogura2005}. Huhne et al.
\cite{Huhne1998} showed that the original approach of Drittler can be used
as well in the relativistic KKR method, where instead of the
Schr\"odinger equation, the Dirac equation has to be solved. One of
the disadvantages of the original implementation is given by the fact
that a certain cut-off radius $R_0$ close to the origin has to be
introduced in order to avoid problems arising from the treatment of the irregular single-site scattering solutions. Recently, it was shown by Zeller \cite{Zeller2014}, that
this approximation might be inconvenient for materials like NiTi and
that the cut-off radius can be chosen arbitrarily small by using an
analytical decoupling scheme and a subinterval procedure with
Chebyshev interpolations in each subinterval.

In the present paper we follow a different approach and discuss
besides the solution via integral equations, the direct solution of the fully
relativistic full-potential single-site scattering problem by means of
standard methods for ordinary differential equations. We
demonstrate that both the regular as well as the irregular scattering
solutions can be obtained with high accuracy. To demonstrate this,
we first discuss the solution of the spherical Coulomb
potential. Furthermore, we derive the asymptotic behaviour of the
relativistic irregular scattering solutions and show that within the
limit towards a non-relativistic description their asymptotic behaviour differs significantly from the non-relativistic irregular scattering solutions. Second, we discuss
the solution of the fully relativistic full-potential single-site
scattering problem for a Mathieu potential in simple cubic lattice and
compare different applied solvers. It will be shown, that linear multi-step methods are a reasonable choice, and therefore,
Adams-Bashforth-Moulton predictor-corrector methods will be used to
calculate the band structure by means of the Bloch spectral function. An
analytically calculated band structure for the Mathieu potential and the
numerically obtained results will be compared.

\section{Single-site scattering\label{single-site}}
In this section we focus on the
Dirac or Kohn-Sham-Dirac equation \cite{MacDonald1979,ramana1979} in atomic Rydberg units ($\hbar=1$, $m=\frac{1}{2}$, $c=\frac{2}{\alpha}\approx 274$) applied to non-magnetic systems,
\begin{equation}
  \left[-i c\, \vec{\mat{\alpha}}\cdot\vec{\nabla} + \frac{1}{2}\mat{\beta} c^2+\mat{I}_4 V_\text{eff}(\vec{r}) \right]\vec{\phi}_\Lambda(\vec{r}) = W\vec{\phi}_\Lambda(\vec{r}),
\end{equation}
where the index $\Lambda$ denotes a combined index $(\kappa,\mu)$.
According to Huhne et al.~\cite{Huhne1998} the equation above can be
solved via expanding the solution $\vec{\phi}_\Lambda(\vec{r})$, which
is a four component spinor-function, into so called spin-angular
functions $\vec{\chi}_{\Lambda'}(\hat{r})$,
\begin{equation}
\vec{\phi}_\Lambda(\vec{r}) = \sum_{\Lambda'} \left(
\begin{array}{c}
g_{\Lambda'\Lambda}(r)\vec{\chi}_{\Lambda'}(\hat{r}) \\ 
i f_{\Lambda'\Lambda}(r)\vec{\chi}_{\overline{\Lambda}'}(\hat{r})
\end{array} 
\right).
\label{ansatz}
\end{equation}
The spin-angular functions, $\vec{\chi}_\Lambda(\hat{r})= \sum_{m_s}
C_\Lambda^{m_s} Y_l^{\mu-m_s}(\hat{r}) \vec{\xi}_{m_s}$, are given by a linear combination of complex spherical harmonics
$Y_l^{m}(\hat{r})$ and the Pauli spinors $\vec{\xi}_{m_s}$, whereas
the expansion coefficients $C_\Lambda^{m_s}$ denote the Clebsch-Gordan
coefficients $C\left(l\, \frac{1}{2}\, j, (\mu-m_s)\, m_s\right)$
\cite{strange1998}. Using the orthogonality of the
spin-angular functions and by defining the matrix $\mat{V}(r)$ with
components $V_{\Lambda \Lambda'} = \left< \vec{\chi}_\Lambda\right|
V_\text{eff} \left|\vec{\chi}_{\Lambda'}\right>$ the following
system of coupled first-order ordinary differential equations can be
obtained:
\begin{align}
-c \left(\frac{\partial}{\partial r}+\frac{1}{r}-\frac{\kappa'}{r} \right) f_{\Lambda'\Lambda} +\frac{c^2}{2} g_{\Lambda'\Lambda} + \sum_{\Lambda''} V_{\Lambda' \Lambda''}\, g_{\Lambda''\Lambda} &= W g_{\Lambda'\Lambda},\label{deq_dif1} \\
 c \left(\frac{\partial}{\partial r}+\frac{1}{r}+\frac{\kappa'}{r} \right) g_{\Lambda'\Lambda}-\frac{c^2}{2} f_{\Lambda'\Lambda} +\sum_{\Lambda''} V_{\overline{\Lambda}'\, \overline{\Lambda}''} f_{\Lambda''\Lambda}   &= W f_{\Lambda'\Lambda}\label{deq_dif2}.
\end{align}
By definition, the potential $V_\text{eff}(\vec{r})$ is constructed such
that it is non-zero within the Wigner-Seitz cell of a certain atomic
site and zero outside. To obtain this behaviour
$V_\text{eff}(\vec{r})$ is given as the product of a non-spherical potential
$U_\text{eff}(\vec{r})$ and the shape-truncation function $\Theta(\vec{r})$
\cite{Stefanou1990,Stefanou1991},
\begin{equation}
 V_\text{eff}(\vec{r}) = U_\text{eff}(\vec{r})\Theta(\vec{r}).
\end{equation}
The elements of the potential matrix of the effective potential $V_\text{eff}(\vec{r})$, based on an expansion in terms of complex spherical harmonics, are given by
\begin{equation}
V_{\Lambda\Lambda'}(r) = \sum_{l''m''}\sum_{m_s} C_\Lambda^{m_s} C_{\Lambda'}^{m_s} V_{l'',m''}(r) G_{l\,l''\,l'}^{(\mu-m_s)\,m''\,(\mu'-m_s)}. 
 \label{relativistic_KKR:potential:vqqz}
\end{equation}
For the construction of the Green function, it is necessary to obtain
two linearly independent solutions $\vec{\phi}_\Lambda(\vec{r})$. First, a solution $\vec{R}_\Lambda(\vec{r})$, which is regular at the origin, and second, a solution $\vec{H}_\Lambda(\vec{r})$, which is
singular at the origin. Outside of the Wigner-Seitz cell both
solutions can be constructed from the spherical Bessel functions
$j_l(r)$ and the spherical Hankel functions $h_l(r)=j_l(r) + i
n_l(r)$, which are a linear combination of the spherical Bessel and the spherical Neumann functions $n_l(r)$. By introducing
the single-site scattering t-matrix $\mat{t}$, the regular single-site
scattering solution outside of the Wigner-Seitz cell can be written in
the following way \cite{Drittler1991b,Huhne1998},
\begin{equation}
 \vec{R}_\Lambda^{\text{outside}}(\vec{r}) = \sum_{\Lambda'} \left[ 
\left(
\begin{array}{c}
j_{l'}(kr) \vec{\chi}_{\Lambda'}(\hat{\vec{r}}) \\ 
e_{\Lambda'} j_{\overline{l}'}(kr)\vec{\chi}_{\Lambda'}(\hat{\vec{r}})
\end{array} 
\right)\,\delta_{\Lambda'\Lambda} - i k 
\left(
\begin{array}{c}
 h_{l'}(kr) \vec{\chi}_{\Lambda'}(\hat{\vec{r}}) \\ 
e_{\Lambda'} h_{\overline{l}'}(kr)\vec{\chi}_{\Lambda'}(\hat{\vec{r}})
\end{array} 
\right) t_{\Lambda'\Lambda}
 \right],
 \label{matching_II}
\end{equation}
by using the abbreviations $e_\Lambda =
\frac{ick}{E+c^{2}}\operatorname{sign}(\kappa)$, $\overline{l} = l -
\operatorname{sign}(\kappa)$, and the relativistic momentum $k= \sqrt{\varepsilon(1+\frac{\varepsilon}{c^2})}$
with $\varepsilon=W-c^2$. The associated irregular
single-site scattering solution is given by
\begin{equation}
 \vec{H}_\Lambda^{\text{outside}}(\vec{r}) =\left(
\begin{array}{c}
h_{l}(kr) \vec{\chi}_{\Lambda}(\hat{\vec{r}}) \\ 
e_\Lambda h_{\overline{l}}(kr)\vec{\chi}_{\Lambda}(\hat{\vec{r}})
\end{array} 
\right).
 \label{matching_II_ir}
\end{equation}
In the following we will discuss two different approaches for the numerical solution
of equation \eqref{deq_dif1} and \eqref{deq_dif2} inside the Wigner-Seitz cell.
\subsection{Numerical solution of the differential equations}
For a numerical treatment of equation \eqref{deq_dif1} and \eqref{deq_dif2}
it is common to transform the large and the small component of
the solution according to
\begin{equation}
 f_{\Lambda'\Lambda}(r) = \frac{Q_{\Lambda'\Lambda}}{cr},\qquad\text{and}\qquad g_{\Lambda'\Lambda}(r) = \frac{P_{\Lambda'\Lambda}}{r}.
\end{equation}
By introducing the matrices $\mat{U}^+(r)= \mat{V}(r)
-\varepsilon\,\mat{I}_4$ and
$\mat{U}^{-}(r)=\left(\frac{\varepsilon}{c^2}+1\right)\,\mat{I}_4
-\frac{1}{c^2}\,\mat{V}(r)$ as well as the spin-orbit coupling matrix
$\mat{K} = \left\{\kappa' \delta_{\Lambda\Lambda'} \right\}$, the
following differential equations in matrix form are obtained,
\begin{align}
&\text{\textbf{I:}} \qquad\,\,  \frac{\mathrm{d}}{\mathrm{d} r} \mat{Q}(r) = \quad\frac{1}{r}\,\mat{K}\cdot\mat{Q}(r) + \mat{U}^+(r)\cdot\mat{P}(r),\label{rad_Ib}\\
&\text{\textbf{II:}}\qquad \frac{\mathrm{d}}{\mathrm{d} r} \mat{P}(r) = -\frac{1}{r}\,\mat{K}\cdot\mat{P}(r) + \mat{U}^-(r)\cdot\mat{Q}(r).\label{rad_IIb}
\end{align}
Employing appropriate initial conditions the coupled equations \eqref{rad_Ib} and \eqref{rad_IIb} can be solved. After this step, the regular scattering solution has to be normalized at the boundary of the Wigner-Seitz cell according to equation
\eqref{matching_II}. The normalized expansion coefficients of
the regular single-site scattering solution inside the Wigner-Seitz
cell $P^{\text{inside}}_{\Lambda'\Lambda}(r)$ can be found as a linear
combination of the numerically obtained solutions $P_{\Lambda'\Lambda}(r)$,
\begin{equation}
 P^{\text{inside}}_{\Lambda'\Lambda}(r) = \sum_{\Lambda''} P_{\Lambda'\Lambda''}(r)\, a_{\Lambda''\Lambda},\quad\text{and}\quad Q^{\text{inside}}_{\Lambda'\Lambda}(r) = \sum_{\Lambda''} Q_{\Lambda'\Lambda''}(r)\, a_{\Lambda''\Lambda}~.
 \label{relativistic_KKR:normalization:e0}
\end{equation}
By introducing the matrices $\mat{X}(r_{\text{BS}})
=\left\{r_{\text{BS}} x_{l}(kr_{\text{BS}}) \delta_{\Lambda\Lambda'} \right\}$ and
$\mat{x}(r_{\text{BS}}) = \left\{r_{\text{BS}} c e_\Lambda
  x_{\overline{l}}(kr_{\text{BS}}) \delta_{\Lambda\Lambda'} \right\}$ ($x = j, h$)
at the radius of the circumscribing sphere $r_{\text{BS}}$ of the
Wigner-Seitz cell, the following algebraic system of linear equations can
be formulated,
\begin{equation}
 \left( 
 \begin{array}{cc}
 \mat{P}(r_{\text{BS}}) & i k \mat{H}(r_{\text{BS}}) \\ 
 \mat{Q}(r_{\text{BS}}) & i k \mat{h}(r_{\text{BS}})
 \end{array} 
 \right)
  \left( 
  \begin{array}{c}
  \mat{a}\\ 
  \mat{t}
  \end{array} 
  \right)
  = 
    \left( 
    \begin{array}{c}
    \mat{J}(r_{\text{BS}})\\ 
    \mat{j}(r_{\text{BS}})
    \end{array} 
    \right).
    \label{relativistic_KKR:normalization:e4}
\end{equation}
To discuss the numerical solution of the ordinary differential equations \eqref{rad_Ib} and \eqref{rad_IIb} we used
{\em Matlab} \cite{matlab} and compared the performance of seven different methods, which
will be introduced briefly in the following. First of all, we used the
standard solvers {\em ode113} and {\em ode15s}, which are an
Adams-Bashforth-Moulton predictor-corrector method of
variable step size and variable order $1,\dots,13$ and an implicit
method of variable step size and variable order $1,\dots,5$ based on
the numerical differentiation formulas \cite{shampine1997,ashino2000}.
Furthermore, the Dormand-Prince method \cite{dormand1980} ({\em ode45})
and the Bogacki-Shampine method \cite{bogacki1989} ({\em ode23}) are
used. Both solver are explicit methods of Runge-Kutta type with variable
step size and orders 5 and 3, respectively. Since the Bulirsch-Stoer
algorithm is used to solve the single-site scattering problem within some KKR codes, we included a Gragg-Bulirsch-Stoer algorithm {\em(GBS)} with both adaptive order
and step size \cite{Fiedler2010}. Last but not least, the fixed step size Runge-Kutta method of order 4 {\em(RK4)} and the fixed step size Adams-Bashforth-Moulton predictor corrector method of order 5 {\em(AB5)} are taken into account, since both are widely applied within the KKR community \cite{zabloudil2005}. For the {\em AB5} method, we use an explicit Adams-Bashforth method of order 5 as predictor which is corrected by an implicit Adams-Moulton method of order 5. The corrector step is repeated $n$ times, until the change of the solution due to the corrector step is smaller than a requested accuracy (PE(CE)$^n$ method). For {\em RK4} and {\em AB5} the ordinary differential equation was transformed analytically by assuming a logarithmic scale $x = \log(r)$. 

\subsection{Numerical solution via integral equations}
Besides a direct solution of the differential equations
\eqref{rad_Ib} and \eqref{rad_IIb} it is possible to solve the
single-site scattering problem by means of integral equations.
Suppose that the single-site scattering solution of the Dirac equation is known for an arbitrary reference system with the potential
$\mathring{V}_\text{eff}(\vec{r})$. These solutions are denoted by
$\mathring{\vec{\Psi}}_\Lambda(\vec{r})$, which could be either regular or irregular single-site scattering solutions. By using the Lippmann-Schwinger equation \cite{Lippmann50},
\begin{equation}
\vec{\Phi}_\Lambda(\vec{r})
= \mathring{\vec{\Psi}}_\Lambda(\vec{r})
+ \int_{\Omega_{WS}} d^{3}r~\mathring{\mat{G}}_{\text{S}}(E; \vec{r}, \vec{r}')\,\Delta V(\vec{r'})\,\vec{\Phi}_\Lambda(\vec{r}')~,
\label{Lippmann_Schwinger}
\end{equation}
it is possible to calculate the solutions $\vec{\Phi}_\Lambda(\vec{r})$ corresponding to the potential
$V_\text{eff}(\vec{r}) = \Delta V(\vec{r}) +
\mathring{V}_\text{eff}(\vec{r})$. The quantity $\mathring{\mat{G}}_{\text{S}}(E; \vec{r}, \vec{r}')$ in equation \eqref{Lippmann_Schwinger} denotes the single-site scattering Green function of the reference system and is constructed from the regular as well as the irregular single-site scattering wave functions
$\mathring{\vec{R}}_{\Lambda}(E,\vec{r})$ and
$\mathring{\vec{H}}_{\Lambda}(E,\vec{r}')$ in the following way:
\begin{equation}
 \mathring{\mat{G}}_{\text{S}}(E; \vec{r}, \vec{r}') = \sum_{\Lambda}\left(\mathring{\vec{H}}_{\Lambda}(E,\vec{r})\mathring{\vec{R}}_{\Lambda}(E,\vec{r}')^\times\theta(r-r') + \mathring{\vec{R}}_{\Lambda}(E,\vec{r})\mathring{\vec{H}}_{\Lambda}(E,\vec{r}')^\times\theta(r-r')\right).
 \label{single_Green}
\end{equation}
The quantities $\mathring{\vec{R}}_{\Lambda}^n(E,\vec{r_n})^\times$ and
$\mathring{\vec{H}}^n_{\Lambda}(E,\vec{r}_n)^\times$ denote the so-called
left-hand side solutions, which were discussed in detail by Tamura
\cite{tamura1992}. By expanding the single-site scattering wave
function in terms of the spin-angular functions \eqref{ansatz} and by
using equation \eqref{single_Green}, the
Lippmann-Schwinger equation \eqref{Lippmann_Schwinger} for the regular
and irregular single-site scattering wave functions can be
reformulated in terms of the following matrix equations,
\begin{equation}
\vec{R}_{\Lambda'\Lambda}(z;r) = \sum_{\Lambda''}\Biggl[
                        \mathring{\vec{R}}_{\Lambda'\Lambda''}(z;r)\,A_{\Lambda'',\Lambda}(z; r)
                      + \mathring{\vec{H}}_{\Lambda'\Lambda''}(z;r)\,B_{\Lambda'',\Lambda}(z; r)\Biggr],
\end{equation}
and
\begin{equation}
    \vec{H}_{\Lambda'\Lambda}(z;r) 
 = \sum_{\Lambda''}
    \Biggl[
    \mathring{\vec{R}}_{\Lambda'\Lambda''}(z;r)\,C_{\Lambda'',\Lambda}(z;r)
  + \mathring{\vec{H}}_{\Lambda'\Lambda''}(z;r)\,D_{\Lambda'',\Lambda}(z;r)
    \Biggr].
\end{equation}
The matrices $\mat{A}(z,r)$, $\mat{B}(z,r)$, $\mat{C}(z,r)$ and
$\mat{D}(z,r)$ contain the integrals over the radial amplitudes $g^{R,H}_{\Lambda'\Lambda}(r)$ and
$f^{R,H}_{\Lambda'\Lambda}(r)$ as well as the functions $x^{R,H}_{\Lambda'\Lambda}(r)$ and
$y^{R,H}_{\Lambda'\Lambda}(r)$, which include the solutions of the
reference system and the change of the potential $\Delta V(\vec{r})$,
\begin{align}
    A_{\Lambda',\Lambda}(z; r)
& = \delta_{\Lambda',\Lambda}\\
& + \sum_{\Lambda_{2}} \int_{r}^{R_{BS}} dr' r'^{2}
    \Biggl(
      x^{H}_{\Lambda',\Lambda_{2}}(z;r')\,g^{R}_{\Lambda_{2},\Lambda}(z; r')
    + y^{H}_{\Lambda',\Lambda_{2}}(z;r')\,f^{R}_{\Lambda_{2},\Lambda}(z; r')
    \Biggr)~,\\
    B_{\Lambda',\Lambda}(z; r)
& = \sum_{\Lambda_{2}} \int_{0}^{r} \quad dr' r'^{2}
    \Biggl(
      x^{R}_{\Lambda',\Lambda_{2}}(z;r')\,g^{R}_{\Lambda_{2},\Lambda}(z; r')
    + y^{R}_{\Lambda',\Lambda_{2}}(z;r')\,f^{R}_{\Lambda_{2},\Lambda}(z; r')
    \Biggr)~,\\
    C_{\Lambda',\Lambda}(z;r)
& = \sum_{\Lambda_{2}}\int_{r}^{R_{BS}} dr' r'^{2}
    \Biggl(
    x^{H}_{\Lambda', \Lambda_{2}}(z;r')\,g^{H}_{\Lambda_{2}, \Lambda}(z;r')
  + y^{H}_{\Lambda', \Lambda_{2}}(z;r')\,f^{H}_{\Lambda_{2}, \Lambda}(z;r')
    \Biggr),\\
D_{\Lambda',\Lambda}(z;r)
& = \delta_{\Lambda',\Lambda}\nonumber\\
& - \sum_{\Lambda_{2}}\int_{r}^{R_{BS}} dr' r'^{2}
    \Biggl(
    x^{R}_{\Lambda', \Lambda_{2}}(z;r')\,g^{H}_{\Lambda_{2}, \Lambda}(z;r')
  + y^{R}_{\Lambda', \Lambda_{2}}(z;r')\,f^{H}_{\Lambda_{2}, \Lambda}(z;r')
    \Biggr)~.
\end{align}
The explicit expressions of $x^{R,H}_{\Lambda'\Lambda}(r)$ and
$y^{R,H}_{\Lambda'\Lambda}(r)$ are given by
\begin{align}
    x^{H}_{\Lambda',\Lambda_{2}}(z;r')
& = -ip \sum_{\Lambda_{1}} \mathring{g}^{H}_{\Lambda_{1},\Lambda'}(z; r')\Delta V^{+}_{\Lambda_{1},\Lambda_{2}}(r')~,\\
    y^{H}_{\Lambda',\Lambda_{2}}(z;r')
& = -ip \sum_{\Lambda_{1}} \mathring{f}^{H}_{\Lambda_{1},\Lambda'}(z; r')\Delta V^{-}_{\Lambda_{1}, \Lambda_{2}}(r')~,\\
    x^{R}_{\Lambda',\Lambda_{2}}(z;r')
& = -ip \sum_{\Lambda_{1}} \mathring{g}^{R}_{\Lambda_{1},\Lambda'}(z; r')\Delta V^{+}_{\Lambda_{1},\Lambda_{2}}(r')~,\\
    y^{R}_{\Lambda',\Lambda_{2}}(z;r')
& = -ip \sum_{\Lambda_{1}}\mathring{f}^{R}_{\Lambda_{1},\Lambda'}(z; r')\Delta V^{-}_{\Lambda_{1}, \Lambda_{2}}(r')~.
\end{align}
In general, the Lippmann-Schwinger equation \eqref{Lippmann_Schwinger}
is solved iteratively in terms of Born series by starting with the
reference solution $\vec{\Phi}^{(0)}_\Lambda(\vec{r})=
\mathring{\vec{\Psi}}_\Lambda(\vec{r})$ on the right hand side of equation \eqref{Lippmann_Schwinger} and by calculating a better approximation $\vec{\Phi}^{(1)}_\Lambda(\vec{r})$ by integration.
This scheme is repeated, until the difference between two successive approximations, $\vec{\Phi}^{(n)}_\Lambda(\vec{r})$ and
$\vec{\Phi}^{(n-1)}_\Lambda(\vec{r})$, is reasonably small.
\section{The Coulomb potential}
\subsection{Asymptotic behaviour}
\begin{figure}
	\centering
	\includegraphics{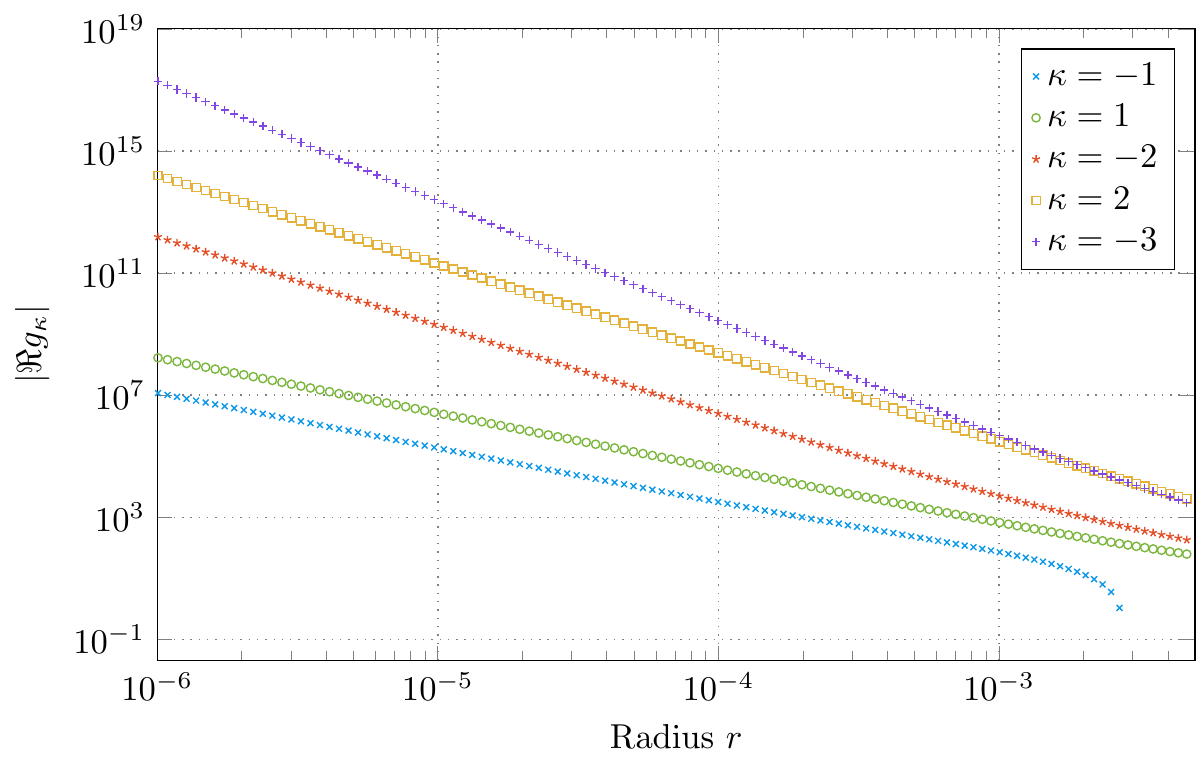}
	\caption{Double-logarithmic plot of the real part of relativistic irregular single-site scattering solutions for a Coulomb potential ($Z=79$) close to the origin. \label{Coulomb_f1}}
\end{figure}

The Coulomb potential is spherically symmetric and hence, the
expansion into spherical harmonics only consists of one (spherically symmetric) component $V_{00}(r)$,
\begin{equation}
 V_{lm}(r) =- \frac{1}{\sqrt{\pi}}\frac{Z}{r}\, \delta_{l,0}\, \delta_{m,0}.
\end{equation}
The associated ordinary differential equations are independent of the relativistic magnetic quantum number $\mu$ and diagonal in $\kappa$, but coupled with respect to $g_\kappa$ and $f_\kappa$,
\begin{align}
 \left[\frac{c^2}{2} - \frac{2 Z}{r} -W \right] g_\kappa(r) + \left[\frac{\kappa c}{r} - \frac{c}{r}\frac{\mathrm{d}}{\mathrm{d}r} r \right] f_\kappa(r) &= 0 \label{coulomb_eq1}\\ 
   \left[\frac{\kappa c}{r} + \frac{c}{r}\frac{\mathrm{d}}{\mathrm{d}r} r \right] g_\kappa(r) - \left[\frac{c^2}{2} + \frac{2 Z}{r} + W \right] f_\kappa(r) &= 0.\label{coulomb_eq2}
\end{align}
It was pointed out by Swainson and Drake \cite{swainson1991I} that
equations \eqref{coulomb_eq1} and \eqref{coulomb_eq2} can be decoupled by
transforming the radial solutions according to
\begin{equation}
 \left(\begin{array}{c}
\tilde{g}_\kappa(r) \\ 
\tilde{f}_\kappa(r)
\end{array}  \right)
=\left(\begin{array}{cc}
1 & X \\ 
X & 1
\end{array}  \right) \cdot
 \left(\begin{array}{c}
g_\kappa(r) \\ 
f_\kappa(r)
\end{array}  \right).
\label{coulomb_eq3}
\end{equation}
Here, the newly introduced quantities are given by
$X = \frac{2(\gamma-\kappa)}{c Z}$ and
$\gamma = \sqrt{\kappa^2 - \frac{4Z^2}{c^2}}$. By using the abbreviations
$\epsilon=\frac{W}{c}$, $\omega^2 = (\frac{c}{2} - \frac{W}{c})^2$,
and by combining equation \eqref{coulomb_eq3} together with \eqref{coulomb_eq1} and
\eqref{coulomb_eq2}, two differential equations of second order can be
obtained, which are similar to the form of the radial Schr\"odinger
equation,
\begin{align}
 \left[\frac{\mathrm{d}^2}{\mathrm{d}r^2} + \frac{2}{r} \frac{\mathrm{d}}{\mathrm{d}r} - \frac{\gamma (\gamma+1)}{r^2}+\frac{2 \alpha Z \epsilon}{r} - \omega^2 \right] \tilde{g}_\kappa(r) &= 0, \label{coulomb_deq1}\\
 \left[\frac{\mathrm{d}^2}{\mathrm{d}r^2} + \frac{2}{r} \frac{\mathrm{d}}{\mathrm{d}r} - \frac{\gamma (\gamma-1)}{r^2}+\frac{2 \alpha Z \epsilon}{r} - \omega^2 \right] \tilde{f}_\kappa(r) &= 0.\label{coulomb_deq2}
\end{align}
Close to the origin ($r\ll1$), in the asymptotic limit, only the angular momentum terms ($\sim
r^{-2}$) have to be taken into account, whereas the potential ($\sim
r^{-1}$) and the constant factor $\omega^2$ can be neglected. The
solution of the resulting differential equations are rational functions of
the following form,
\begin{align}
 \tilde{g}_\kappa &= c_1 r^{\gamma} + c_2 r^{-\gamma-1},  \label{coulomb_asym1}\\
 \tilde{f}_\kappa &= c_3 r^{\gamma-1} + c_4 r^{-\gamma}. \label{coulomb_asym2}
\end{align}
Since both solutions $g_\kappa$ and $f_\kappa$ are linear combinations
of $\tilde{g}_\kappa$ and $\tilde{f}_\kappa$, it can be verified, that
the leading term for the irregular solutions is given by $r^{-\gamma-1}$. In the non-relativistic limit ($\frac{4 Z^2}{c^2} \approx 0$) the asymptotic behaviour of an
irregular $s$-wave function ($\kappa=-1$, $l=0$) close to the origin
is given by $r^{-\left|\kappa\right|-1}=r^{-2}$ which is in
contradiction to the non-relativistic solution $r^{-l-1}=r^{-1}$.
Since the relativistic quantum number $\kappa$ takes on values either
$\kappa=l$ or $\kappa=-l-1$, this disagreement can be generalized for
all irregular wave functions with $\kappa=-l-1$. To verify the
solution behaviour for $r\ll1$, a double logarithmic plot of the
numerical solution for $Z=79$ and $c=\frac{2}{\alpha}$ and various
values for $\kappa$ is illustrated in figure \ref{Coulomb_f1}. The
predicted asymptotic behaviour is clearly revealed. 

\subsection{Numerical accuracy\label{Coulomb_accuracy}}
For the discussion of the numerical accuracy for the solution of
the differential equation (equations \eqref{coulomb_deq1} and \eqref{coulomb_deq2}), a Coulomb potential 
with an atomic number of $Z=79$ was used since it represents the
element gold (Au). Due to the large atomic mass and non-negligible spin-orbit coupling, which causes the typical golden colour, it is a prominent example for relativistic effects. Solutions were obtained up to a maximal angular momentum quantum number of $l=5$. To obtain a scattering state, the energy of the scattering solution was chosen to be $\varepsilon = 1$. In the following test, a spherical cell was constructed including a minimal radius of $r_0=10^{-4}$ and a maximal radius of $r_{\text{BS}}=3$.

\begin{figure}
  \centering
  \includegraphics[width=\textwidth]{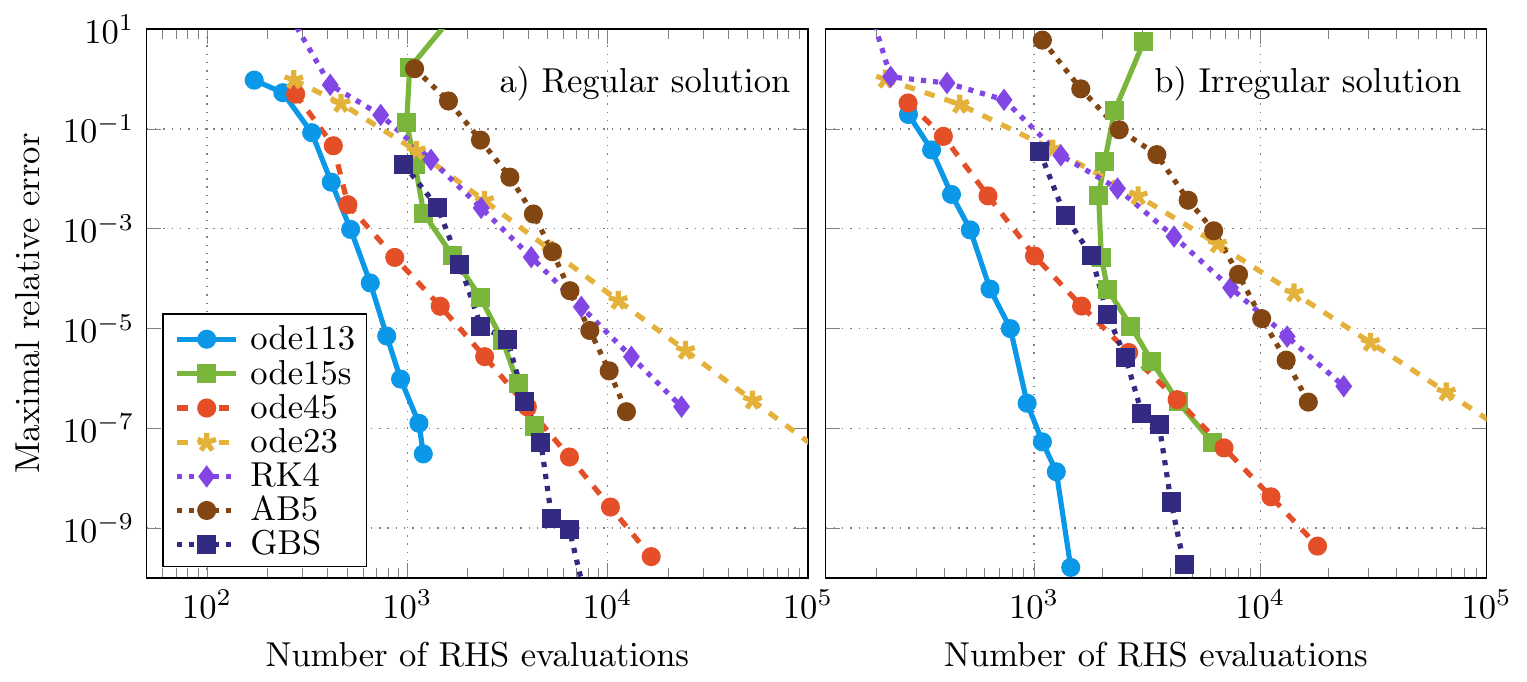}
  \caption{Maximal relative error versus number of right-hand
    side (RHS) evaluations of the Dirac equation for a Coulomb potential using different methods for the solution.\label{coulomb:accuracy:f1}}
\end{figure}

\begin{figure}[t!]
  \centering \subfloat[Solution on a radial
  mesh\label{coulomb:accuracy:f2a}]{
    \includegraphics[width=0.49\textwidth]{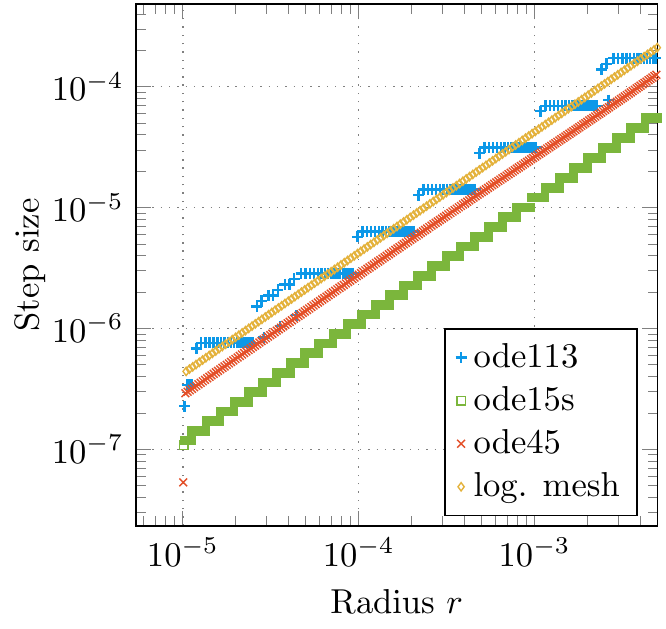}}
  \subfloat[Solution using a logarithmic
  scale\label{coulomb:accuracy:f2b}]{
    \includegraphics[width=0.49\textwidth]{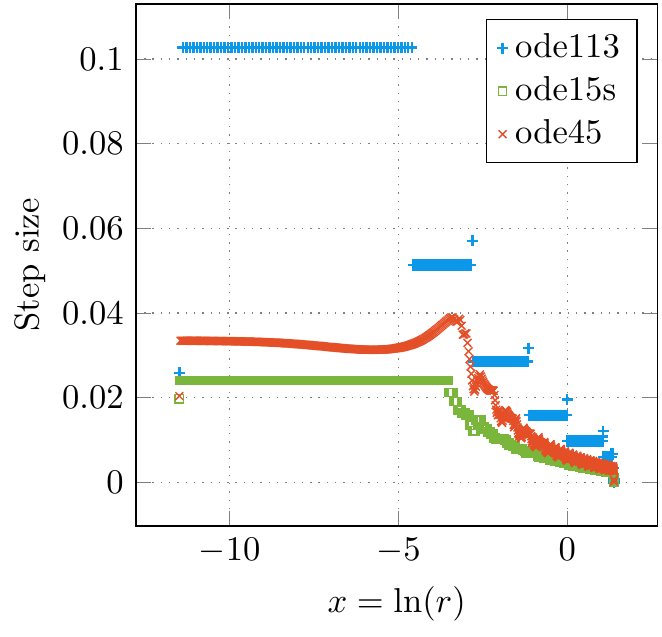}}\hfill
  \caption{Step size for the solution of the Coulomb-Dirac problem
    using adaptive methods. \label{coulomb:accuracy:f2}}
\end{figure}

The numerical solutions obtained by using different solvers were compared with a reference solution, which was obtained by using
{\em ode113} and very high accuracy goals. For the solvers taken from
the {\em Matlab ode-suite} absolute and relative tolerances were chosen to
be equal and values between $10^{-1}$ and $10^{-10}$ were used. The maximal relative error of the numerically obtained solutions as a function of the number of right-hand side evaluations of the differential equation is shown in figure \ref{coulomb:accuracy:f1}.
First of all, it can be verified that the numerical accuracy of the solution
of the regular single-site scattering solution (figure \ref{coulomb:accuracy:f2a}) is similar to the
results of the irregular single-site scattering solution (figure \ref{coulomb:accuracy:f2b}). Hence, the
following statements can be given for both types of solutions. As summarized by S\"oderlind et al. \cite{soderlind2014} an ordinary differential equation is called stiff, if the numerical solution via implicit methods performs much better than the solution via explicit methods. However, for the present example it can be verified that the method {\em ode113}, which is a method for non-stiff equations, performs better
than the implicit method {\em ode15s}. Since we observed poor performance for other implicit methods, we conclude that the underlying differential equations for the Coulomb potential are non-stiff. The performance of the Dormand-Prince method ({\em ode45}) is very good for crude tolerances and becomes comparable to {\em ode15s} for fine tolerances.  The Adams-Bashforth-Moulton predictor-corrector method of order 5 with fixed step size ({\em AB5}) is rather expensive for high tolerances. But, due to the higher order, the performance is better in comparison to the methods of the Runge-Kutta type {\em ode23} and {\em RK4}, if high accuracy goals are demanded.  Also in comparison to {\em ode45}, which has the same order, less evaluations of the right-hand side of the differential equation for high accuracy goals are necessary; it is therefore more efficient. The implemented Gragg-Bulirsch-Stoer method with both adaptive order and step-size ({\em GBS}) \cite{Fiedler2010} is able to solve the differential equations with very high orders. But, for the accuracy goals in practice ($\approx 10^{-8}$) it needs about five times as many evaluations of the right hand side of the differential equation in comparison to {\em ode113}.

In many implementations of the KKR method, solvers with fixed step
size are used for spherically symmetric atomic potentials, whereas a logarithmic
mesh of type $x = \log(r)$ or similar is employed \cite{zabloudil2005}. The reason for this is that the wave functions are highly oscillating close to the
nucleus, and are smooth for larger values of $r$. To verify that
a logarithmic mesh is a reasonable choice, the step size for
adaptive methods close the origin was investigated (see figures \ref{coulomb:accuracy:f2a} and \ref{coulomb:accuracy:f2b}). The step size during the numerical solution of the Coulomb-Dirac problem using
{\em ode113}, {\em ode15s}, and {\em ode45} is illustrated in figure \ref{coulomb:accuracy:f2a} and compared with the step size of a logarithmic mesh. It can be verified that the step size used by {\em ode45} is similar to the characteristic of a logarithmic mesh. However, the stair-case-like behaviour of {\em ode113} and {\em ode15s} occurs due to the step-size strategy of the method itself, i.e. the change of the step-size is avoided as much as possible \cite{shampine1997}. Analogously, it is possible to transform the differential equations \eqref{coulomb_deq1} and \eqref{coulomb_deq2} to a logarithmic scale $x=\log(r)$ and to investigate the varying step size of the methods {\em ode113}, {\em ode15s} and {\em ode45} for the numerical solution of the transformed equations (see figure \ref{coulomb:accuracy:f2b}). It can be verified, that all three solvers adopt a constant step size close to the origin ($x < -5$), which again reassures the choice of a
logarithmic scale for methods with fixed step size.

\section{The Mathieu potential}
\subsection{Evaluation of the potential}
A suitable test system for our full-potential implementation is given by a Mathieu potential \cite{Yeh1990},
\begin{equation}
V(\vec{r}) = -U_0\left(\cos(a x) + \cos(a y) + \cos(a z) \right)~,
\label{mathieu:potential:eq1}
\end{equation}
that is periodic and represents here a simple cubic crystal structure. Moreover, it is highly anisotropic with respect to different crystallographic directions and thus, it is ideal to serve as a test system. For our method it is necessary to express equations \eqref{mathieu:potential:eq1} in terms of complex spherical harmonics (see section \ref{single-site}). By using the exponential form of the cosine function and by applying Bauer's theorem, 
\begin{equation}
 e^{i \vec{k}\cdot\vec{r}} = 4 \pi \sum_{l=0}^{\infty}\sum_{m=-l}^l i^l j_l(kr)Y_l^m(\hat{\vec{r}})Y_l^{m}(\hat{\vec{k}})^*~,
 \label{mathieu:potential:eq3}
\end{equation}
we end up with the following equation,
\begin{equation}
 V(\vec{r}) = - \sqrt{4\pi} U_0 \sum_{l=0}^{\infty}\sum_{m=-l}^l \cos\left(l\frac{\pi}{2}\right) \sqrt{2l+1} \left[
\delta_{m,0} +  f_{lm} \right] j_l(kr) Y_l^m(\hat{\vec{r}}),
\label{Mathieu:expansion:e1}
\end{equation}
with coefficients $f_{lm}$ given by
\begin{equation}
 f_{lm} = \left\{
 \begin{array}{cl}
  \left((-1)^\frac{m}{2}+1\right)(-1)^{\frac{l}{2}} \sqrt{ \frac{(l+m-1)!!}{(l+m)!!}\frac{(l-m-1)!!}{(l-m)!!}} & m\text{~is even} \\ 
  0 & m\text{~is odd}
  \end{array}  
   \right. .
\end{equation}
\subsection{Numerical Accuracy}
Analogously to section \ref{Coulomb_accuracy}, where the numerical
accuracy was discussed for the Coulomb potential, our numerical test environment was used to solve the differential equations \eqref{rad_Ib} and \eqref{rad_IIb} for the Mathieu potential. According to Yeh et al. \cite{Yeh1990}, the lattice constant was chosen to be $a=2\pi$ and the pre-factor $U_0$ was set to $U_0=-0.5$. The expansion of the solution in terms of spin-angular functions was evaluated up to a maximal angular momentum quantum number of $l_{\text{max}}=5$. Due to the high value of $l_{\text{max}}$ the matrices $\mat{P}$ and $\mat{Q}$ in \eqref{deq_dif1} and \eqref{deq_dif2} are of dimension $72 \times 72$.

\begin{figure}[t!]
\includegraphics[width=\textwidth]{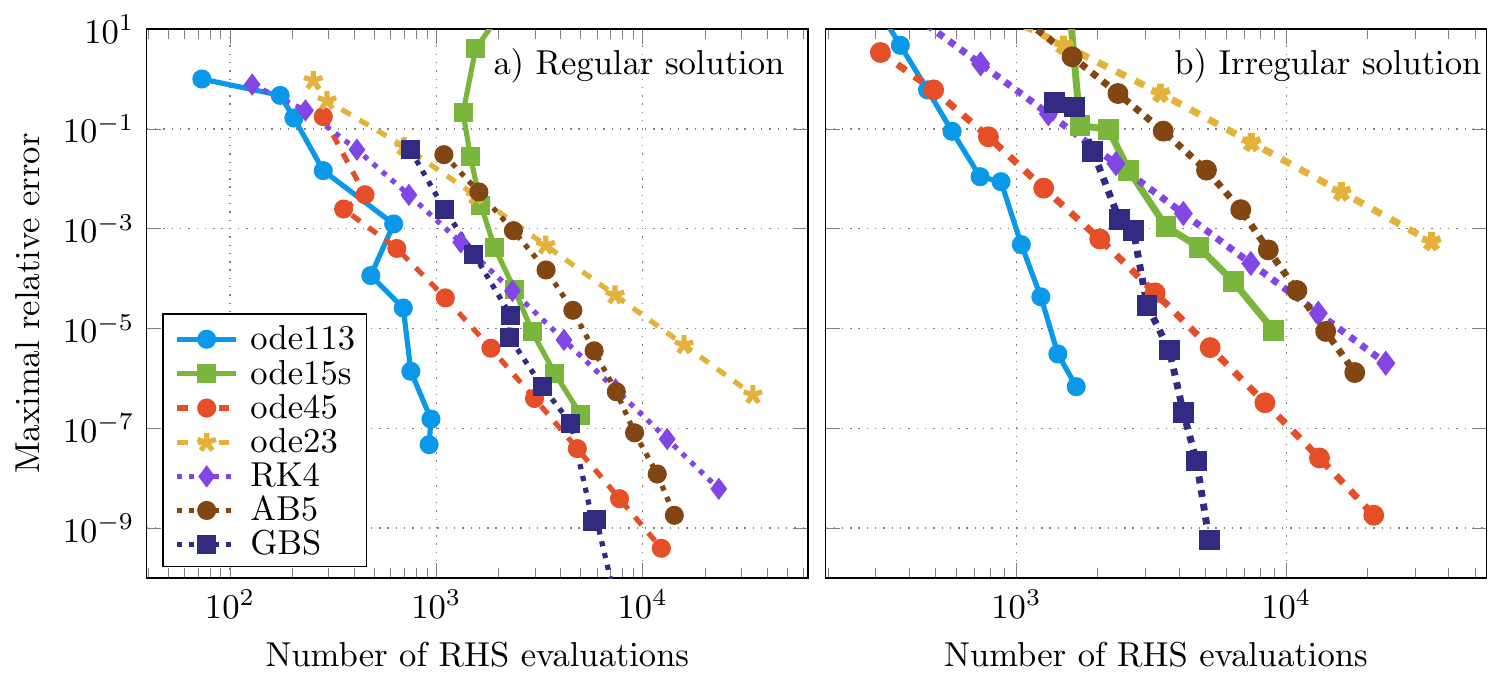}
\caption{Maximal relative error versus number of right-hand
  side (RHS) evaluations of the Dirac equation for a Mathieu potential using
  different methods for the solution.\label{mathieu:accuracy:f1}}
\end{figure}

The maximal relative error between a very precise reference solution and the numerically obtained solution of different solvers versus the number of evaluations of the right-hand side of the differential equations are
shown in figure \ref{mathieu:accuracy:f1}. The general behaviour using different methods is similar for regular and irregular single-site scattering solutions. It can be verified that for a particular method and for the same number of evaluations of the right-hand side of the differential equation the maximal relative error for the regular solution is by approximately 2 orders of magnitude smaller compared to the irregular solution. This is due to very large absolute values of the irregular single-site scattering solutions close to the origin. As shown in the example for the Coulomb potential, we conclude that the underlying differential equations for the Mathieu potential can be regarded as non-stiff, since the performance of the method {\em ode113} is much better than the performance of the method {\em ode15s} \cite{soderlind2014}. Again, the performance of the Adams-Bashforth-Moulton
predictor-corrector method of order 5 ({\em AB5}) is worse than the explicit
Runge-Kutta methods {\em RK4} and {\em ode45} for crude tolerances. However due to the higher order it is more efficient if high accuracy goals are required. Especially for the irregular solutions {\em ode23} fails to give a reasonable solution for an appropriate amount of evaluations of the right-hand side of the differential equations.

The differential equations \eqref{rad_Ib} and \eqref{rad_IIb} are characterized by the effective potential and since the Mathieu potential is analytically known, methods with adaptive step-size like {\em ode113} are a reasonable choice. In general, the effective potential within each iteration of the KKR method is given on a discrete mesh  and, hence, a method with fixed step size is appropriate, since interpolations of the potential between mesh points are avoided. For the Mathieu potential, it can be verified (see figure \ref{mathieu:accuracy:f1}) that the numerical solution of the full-potential single-site scattering problem can be obtained by linear multi-step methods for non-stiff equations, e.g. by applying an Adams-Bashforth-Moulton predictor-corrector method. Therefore, the method {\em AB5} was implemented within the computer code {\em Hutsepot} \cite{luders2001}. 

\begin{figure}[t!]
\centering
\includegraphics[]{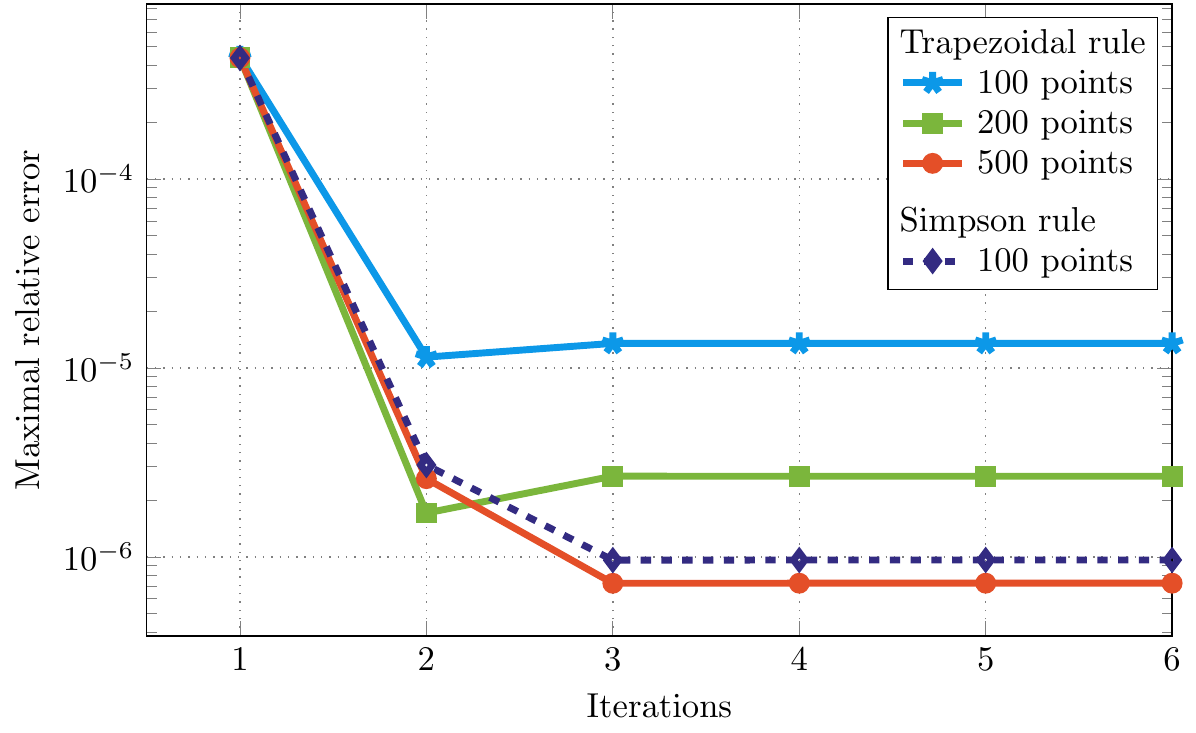}
\caption{Maximal relative error versus number of iterations within the
  Born series for the solution of the Dirac equation for an Mathieu
  potential by means of integral
  equations.\label{mathieu:accuracy:f2}}
\end{figure}

Besides the direct solution of the differential equation, the solution of the
Dirac equation for a Mathieu potential was investigated by means of
integral equations. For the integration, the trapezoidal rule as well
as the Simpson rule was used. In figure \ref{mathieu:accuracy:f2} the
maximal relative error of the solution is plotted versus the number of
iterations within the Born series for the solution of the
Lippmann-Schwinger equation for an irregular single-site scattering
solution. It can be verified that the maximal relative error saturates quickly and the Born series converges after three iterations. This is in perfect agreement with the observation of Drittler \cite{Drittler1991a} for the non-relativistic method. Since the deviation from the exact solution is dominated by the error of the numerical quadrature, the accuracy of the solution can be improved by either increasing the number of mesh points for the quadrature or by improving the integration scheme, e.g. using the Simpson rule instead of the trapezoidal rule. In this way the maximal relative error can be decreased by about one order of magnitude for the same number of points, where the integrand is evaluated.

\subsection{Band structure}
After the solution of the Dirac equation at each atomic site $n$, the
regular and the irregular single-site scattering solutions
$\vec{R}^n_{\Lambda}(E,\vec{r}_n)$ and
$\vec{H}^n_{\Lambda}(E,\vec{r}_n)$ can be used to construct the
relativistic multiple-scattering Green function,
\begin{equation}
   \mat{G}(E; \vec{r}_n+\vec{R}_n,\vec{r}_m+\vec{R}_m)
 = \sum_{\Lambda,\Lambda'}\vec{R}_{\Lambda}^n(E,\vec{r_n})
                          \mat{g}_{\Lambda\Lambda'}^{nm}(E)
                          \vec{R}_{\Lambda'}^m(E,\vec{r_m})^\times
 - \delta_{nm} \mat{G}_{\text{S}}(E; \vec{r}_n, \vec{r}'_m). 
 \label{greenfun}
\end{equation}
Analogously to the single-site scattering Green function
\eqref{single_Green}, $\vec{R}_{\Lambda}^n(E,\vec{r}_n)^\times$ and
$\vec{H}^n_{\Lambda}(E,\vec{r}_n)^\times$ denote the left-hand side
solutions \cite{tamura1992}. The matrix $\mat{g}_{\Lambda\Lambda'}^{nm}(E)$ is called structural Green
function matrix and can be obtained from the structure constants
$\mat{G^0}(E)$ of the free space and the single-site scattering t-matrices $\mat{T}(E)=\left\{\mat{t}^n(E)\right\}$ at each site $n$,
\begin{equation}
 \mat{g}(E) = \mat{G}^0(E)\left[\mat{I}-\mat{T}(E)\,\mat{G}^0(E)\right]^{-1}.
 \label{NRKKR_main_eq3}
\end{equation}
\begin{figure}[t!]
\subfloat[$l_{\text{max}}=2$]{\includegraphics[width=0.49\textwidth]{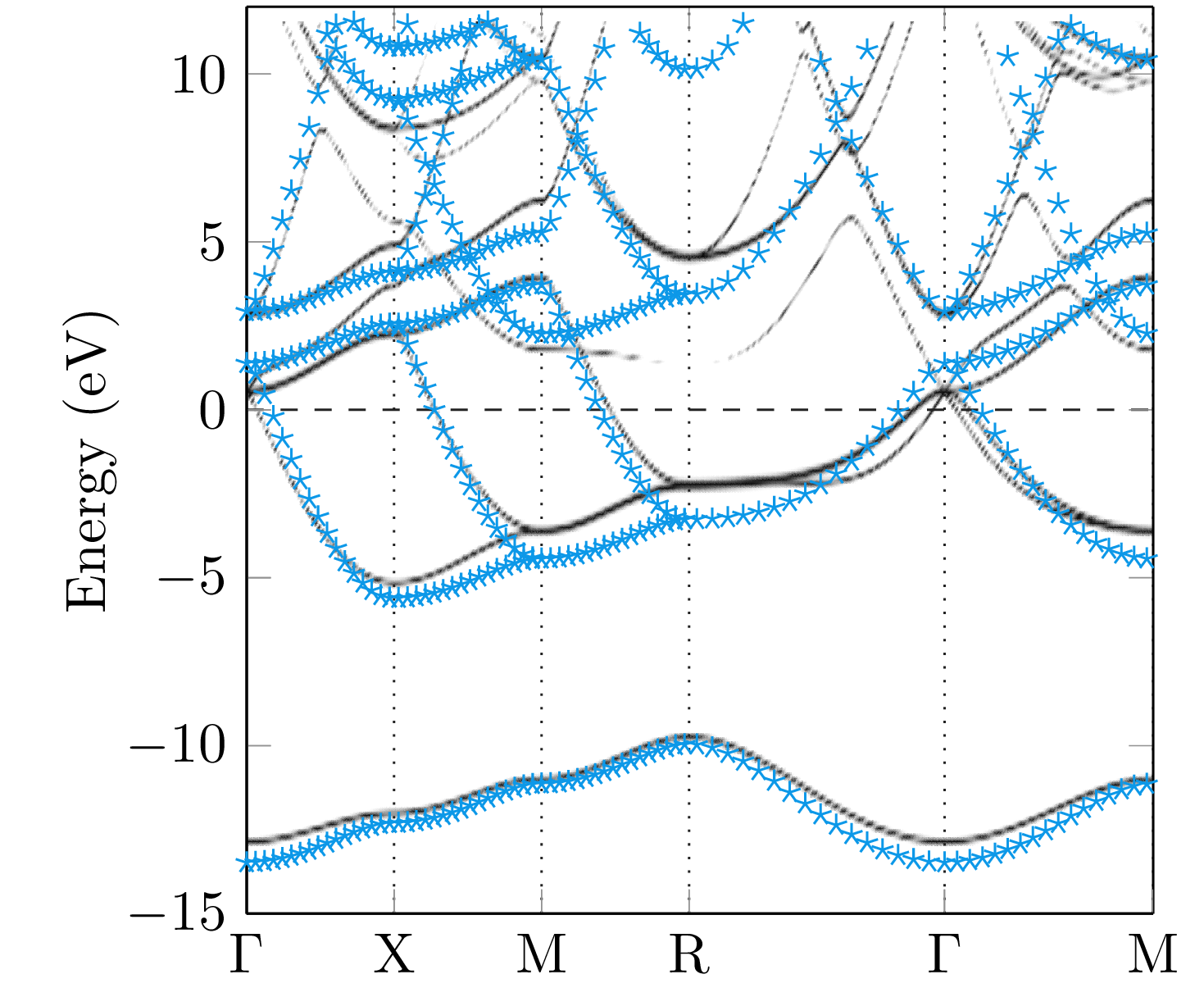}}\hfill
\subfloat[$l_{\text{max}}=3$]{\includegraphics[width=0.49\textwidth]{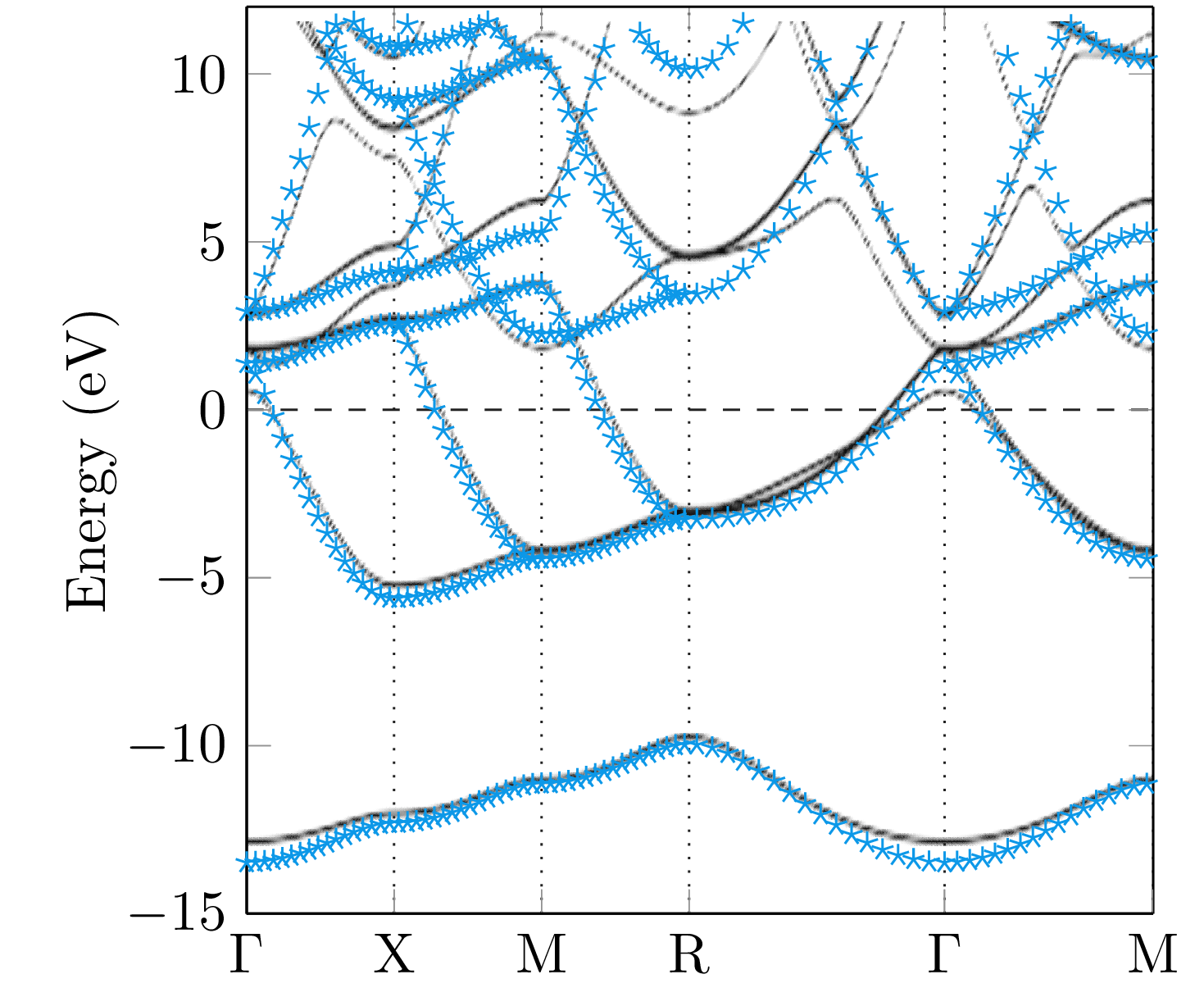}}\\
\subfloat[$l_{\text{max}}=4$]{\includegraphics[width=0.49\textwidth]{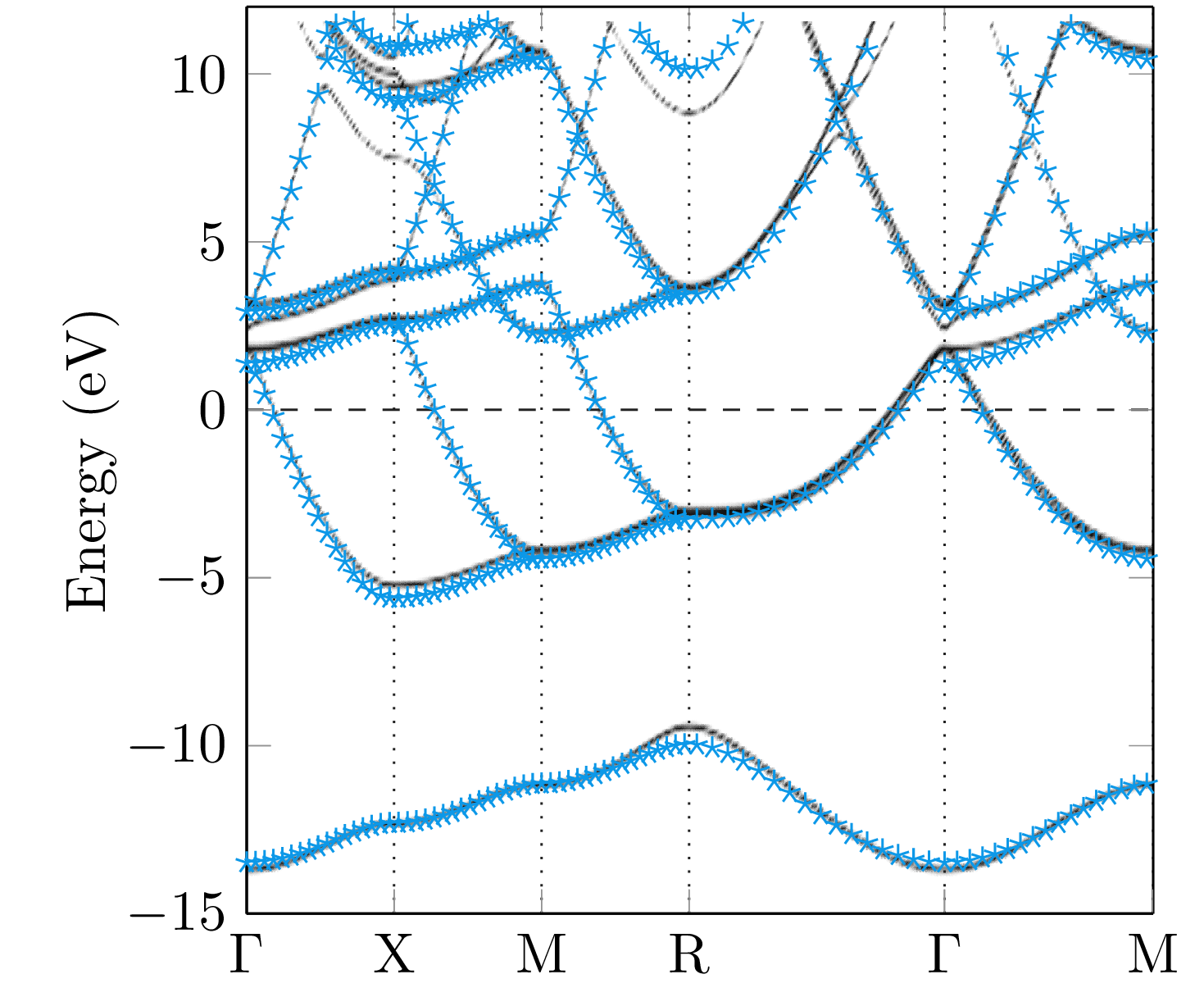}}\hfill
\subfloat[Brillouin zone]{\includegraphics[]{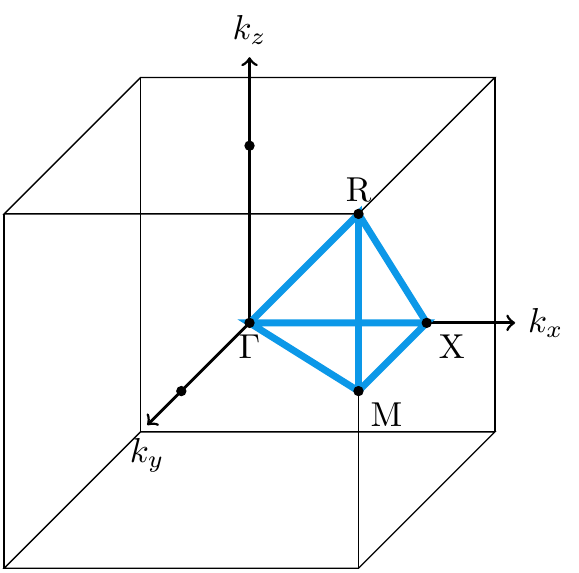}}
\caption{Comparison of the relativistic Bloch spectral function calculated with {\em Hutsepot} (black) and a band structure calculated by means of a plane-wave approach (blue) for the Mathieu potential.\label{mathieu:bands}}
\end{figure}
Since the Green function has a singularity if the energy hits an eigenvalue of the Dirac Hamiltonian, the band structure of a system can be calculated by means of the Bloch spectral function, which is given by the imaginary part of the $\vec{k}$-resolved multiple-scattering Green function \cite{weinberger1990},  
\begin{equation}
A_B(\vec{k},W) = -\frac{1}{\pi} \operatorname{Im} \operatorname{Tr} \left[\sum_{\vec{R}_j\in L} e^{i \vec{k}\cdot \vec{R}_j}\int \mathrm{d}^3 r_j\, \mat{G}(\vec{r}_j,\vec{r}_j+\vec{R}_j,W)  \right].
\label{NRKKR:BlochSF}
\end{equation}
The summation in the last expression belongs to all lattice vectors $\vec{R}_j$ of the lattice denoted by $L$. The trace is a summation of the diagonal components of the $4\times 4$ dimensional Green function. The great advantage of the Mathieu potential is given by the fact, that the band structure can be calculated analytically and, therefore,
it is a good test system for our numerical implementation. In
general, a large number of terms are necessary within the angular momentum expansion of the Green function to obtain the correct energy bands. For illustrative purposes, the Bloch spectral functions obtained for different maximal angular momentum ($l_{\text{max}}=2, 3, 4$) within the expansion \eqref{greenfun} together with the analytically obtained result are compared in Fig.~\ref{mathieu:bands}. As can be seen, high values of $l_{\text{max}}$ are necessary to achieve a reasonable agreement between the band structures. However, the numerical result for $l_{\text{max}}=4$ reflects the general behaviour of the analytical band structure quite nicely.

\section{Conclusion}
An elaborate discussion of the numerical accuracy for the solution of
the spherical Coulomb potential and the Mathieu potential is presented. The solutions were obtained, first, by various standard methods for the solution of ordinary differential equations and, second, by means of an iterative solution of the associated Lippmann-Schwinger equation. From the performance of the solvers for stiff and non-stiff problems we conclude that the differential equations are non-stiff for both systems. The numerical solution can be obtained by using linear multi-step methods like the Adams-Bashforth-Moulton predictor corrector method. For the Coulomb potential the asymptotic behaviour close to the origin ($r\ll 1$) was investigated and it could be shown, that the non-relativistic limit of the irregular single-site scattering solution shows a different asymptotic behaviour in comparison to the associated non-relativistic irregular single-site scattering solution for the case $\kappa = -l-1$.

\section*{Acknowledgements}
We are grateful for financial support by Deutsche Forschungsgemeinschaft in the framework of SFB762 ``Functionality of Oxide Interfaces''. Furthermore, we acknowledge helpful discussions with Dr. Rudolf Zeller, Dr. J\"urgen Henk, Dr. Helmut Podhaisky and Prof. R\"udiger Weiner.
\section*{References}
\bibliographystyle{unsrt}
\bibliography{./IOPLaTeX}

\end{document}